\documentclass[review]{elsarticle}

\usepackage{lineno,hyperref}
\usepackage{array, multirow}
\usepackage{amssymb}
\usepackage[cmex10]{amsmath}
\usepackage{algorithmic}
\usepackage{subcaption}
\usepackage{graphicx}
\usepackage{epstopdf}
\usepackage{subcaption}

\journal{Simulation Modelling Practice and Theory}

\begin{document}

\begin{frontmatter}

\title{A Queuing Model for CPU Functional Unit and Issue Queue Configuration}

\author{S. Carroll\corref{mycorrespondingauthor}}
\cortext[mycorrespondingauthor]{Corresponding author}
\ead{shane.carroll@utsa.edu}
\author{W. Lin\corref{}}
\ead{weiming.lin@utsa.edu}


%\author[mysecondaryaddress]{\corref{mycorrespondingauthor}}

\address{The University of Texas at San Antonio, Department of Electrical and Computer Engineering, San Antonio, TX 78249}


\begin{abstract}
In a superscalar processor, instructions of various types flow through an execution pipeline, traversing hardware resources which are mostly shared among many different instruction types. A notable exception to shared pipeline resources is the collection of functional units, the hardware that performs specific computations. In a trade-off of cost versus performance, a pipeline designer must decide how many of each type of functional unit to place in a processor's pipeline.  In this paper, we model a superscalar processor's issue queue and functional units as a novel queuing network. We treat the issue queue as a finite-sized waiting area and the functional units as servers. In addition to common queuing problems, customers of the network share the queue but wait for specific servers to become ready (e.g., addition instructions wait for adders). Furthermore, the customers in this queue are not necessary ready for service, since instructions may be waiting for operands. In this paper we model a novel queuing network that provides a solution to the expected queue length of each type of instruction. This network and its solution can also be generalized to other problems, notably other resource-allocation issues that arise in superscalar pipelines.
\end{abstract}

\begin{keyword}
	Modeling of computer architecture\sep processor architectures\sep hardware architecture
\end{keyword}

\end{frontmatter}

\section{Introduction}
\label{sec:intro}
\begin{figure}[t]
	\centering
	\includegraphics[width=0.8\textwidth]{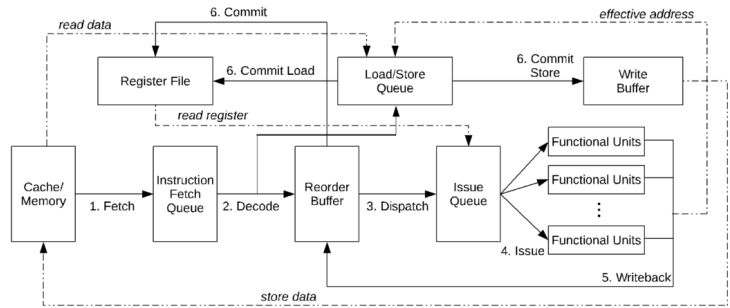}
	\caption{A typical 6-stage RISC instruction pipeline. Pipeline stages are numbered and shown with solid lines; data paths are dashed.}
	\label{fig:pipeline}
\end{figure}

In a superscalar processor, instructions of
various types flow in parallel through a pipeline of several
stages, such as the simple pipeline shown in
Figure~\ref{fig:pipeline}. Instructions start in memory, from
where they are fetched, and traverse the pipeline until their
operations are completed and their results finalized. Instructions
come in various types, such as integer addition, integer
multiplication, floating-point addition, load, store, etc. In most
stages of the pipeline, all types of instructions share a common
space and do not travel along a type-specific data path.

However, as observable in Figure~\ref{fig:pipeline}, there is one
portion of the pipeline in which each specific type of instruction
must travel a type-specific path into their respective functional
units (FU), where the operation (addition, multiplication, etc.)
is completed. At this stage of the pipeline, an issue queue (IQ) acts as a buffer and holds
instructions until they can be issued to an FU. To be ready for
issuing, an instruction must meet the following criteria:
\begin{enumerate}
	\item each of this instruction's operands is ready;
	\item there is an available FU of this instruction's type.
\end{enumerate}
If either of these criteria is not met in a particular clock
cycle, the instruction shall remain in the IQ indefinitely. Given that modern CPUs exploit
instruction-level parallelism, there are typically several of each
type of FU to serve multiple instructions of a single type at a
time. Clearly, in a system with bandwidth $B$ between the IQ and
the FUs, it would be optimal to have the number of each type of FU
equal to $B$ to fulfill its potential. In such a configuration,
criteria (2) is always satisfied and we should intuitively expect
higher throughput than a configuration with any fewer FUs.
However, FUs are expensive in many dimensions, including economic
cost, chip space, and idle power consumption. So by configuring a
system to have ample FUs of each type, we optimize system
performance with respect to this parameter but accept a serious
tradeoff in cost. On the other hand, using just one FU for each
instruction type to minimize costs instead may easily waste some
instruction-level parallelism.

To overcome this problem, an FU configuration must be chosen to
exploit instruction-level parallelism available while maintaining
a reasonable cost. Choosing the best combination of FUs is
inherently difficult due to the size of the respective state
space. To analyze a possible FU configuration for a particular
system, extensive simulation runs have to be performed to
determine the performance implications of the chosen
configuration. In this paper, we take a novel approach to
analyzing an FU configuration by deriving a mathematical model to
predict the expected number of instructions of each type that are
queued in the pipeline stage immediately preceding the FUs.

Note that, in order to maximize program execution throughput, each
type of instruction should maintain a similar flow through the
IQ-FU stage.  Instead, if one type of instruction is consumed by
its FUs slower than others, disregarding the effect of data
dependency among instructions, it most likely results from the
insufficient number of FUs for this type of instruction.
Consequently, more instructions of this type will tend to
overwhelm and clog the IQ, and subsequently slow down the
processing of other instructions as well due to data dependency
among them.  That is, the best FU configuration is the one that
minimizes the total cost of the FU configuration and the queue length of each instruction type.

The goal of this paper is to model the IQ-FU stage as a novel
queuing network and determine, given an FU configuration, derive the
expected queue length of each instruction type.
Under the general hypothesis of queuing theory, the number of
instructions of a given type sitting in IQ can be determined if
the following parameters are given:
\begin{itemize}
	\item the number of FUs for this type (and its execution/input latency),
	\item the prevalence of this type of instructions (i.e. its occurrence
	frequency), and
	\item how likely an instruction of this type in IQ is operand-ready to be issued.
\end{itemize}
Clearly the first parameter is readily available, given as an input
parameter for the network. Each of the other two instead is
real-time program execution dependent, acquisition of which can be either empirically derived or inferred based on system configuration.

\begin{figure}
	\centering
	\begin{subfigure}{0.3\textwidth}
		%\vspace{1.6em}
		\includegraphics[width=\textwidth]{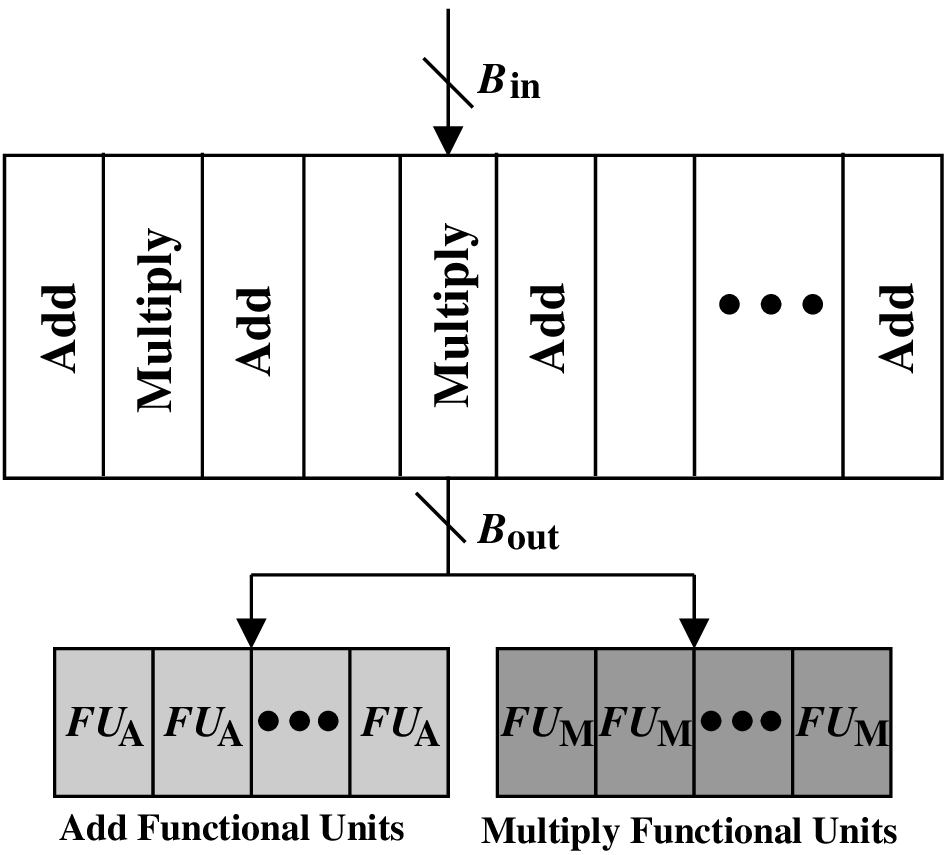}
		\caption{Standard IQ}
		\label{fig:iq_intro:basic}
	\end{subfigure}
	\qquad
	\begin{subfigure}{0.3\textwidth}
		\includegraphics[width=\textwidth]{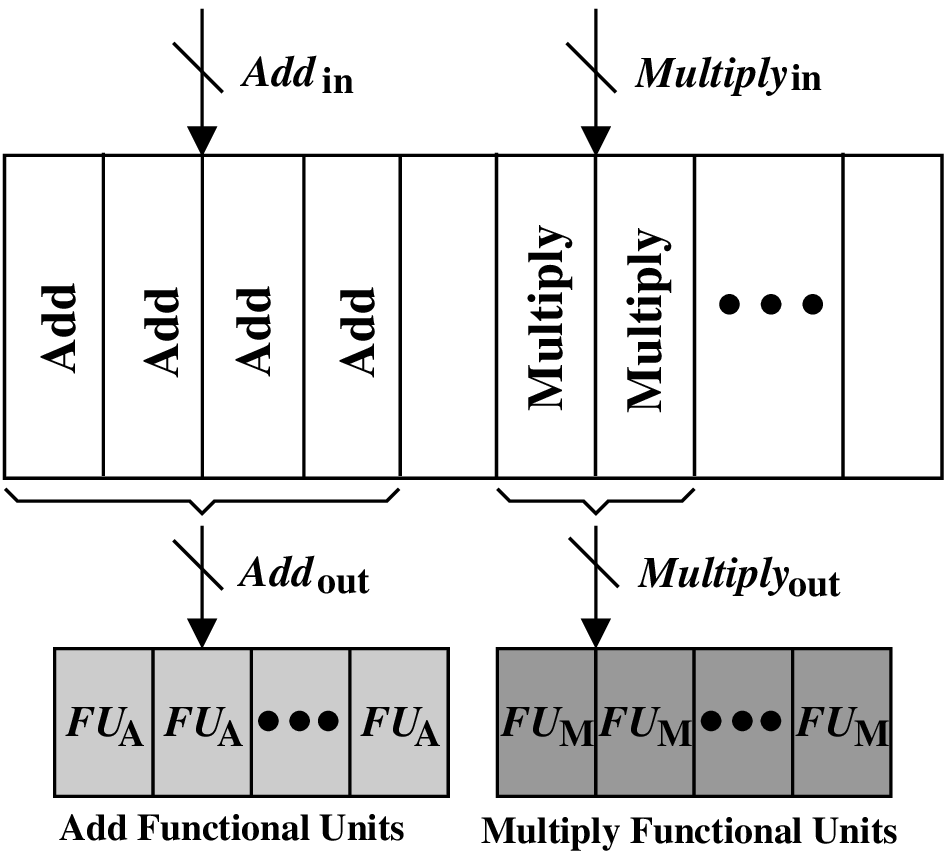}
		\caption{Abstracted IQ}
		\label{fig:iq_intro:equiv}
	\end{subfigure}
	\caption{Abstraction of an IQ, from the standard representation to a logical rearrangement of slots and bandwidth}
\end{figure}

As an illustrating example, a system of two instruction types is
first considered, addition and multiplication instruction types.
Such an IQ may at any time have any combination of addition and
multiplication instructions in it awaiting issue. Of course, the
total number of instructions inside the IQ cannot exceed the IQ's
capacity, which we shall denote as $N$. One example situation of
the aforementioned IQ is shown in Figure~\ref{fig:iq_intro:basic}.

This illustrates a situation in which there are four addition and
two multiplication instructions awaiting issue in the IQ.
Instructions arrive along the incoming bus, which has bandwidth
$B_{in}$, and are issued out of the IQ along the outgoing bus,
which has bandwidth $B_{out}$. Note that we can abstract this IQ
into a logically-equivalent model as shown in
Figure~\ref{fig:iq_intro:equiv}.

Here, we logically group similar instructions and depict the
bandwidth as being split among the instruction types. If we apply
to Figure~\ref{fig:iq_intro:equiv} the constraints that $\mbox{\it
	Add}_{in} + \mbox{\it Multiply}_{in} \le B_{in}$ and $\mbox{\it
	Add}_{out} + \mbox{\it Multiply}_{out} \le B_{out}$ in any
particular clock cycle, we derive an equivalent model as in
Figure~\ref{fig:iq_intro:basic}.

\begin{figure}
	\centering
	\begin{subfigure}{0.4\textwidth}
		\centering
		\includegraphics[width=0.7\textwidth]{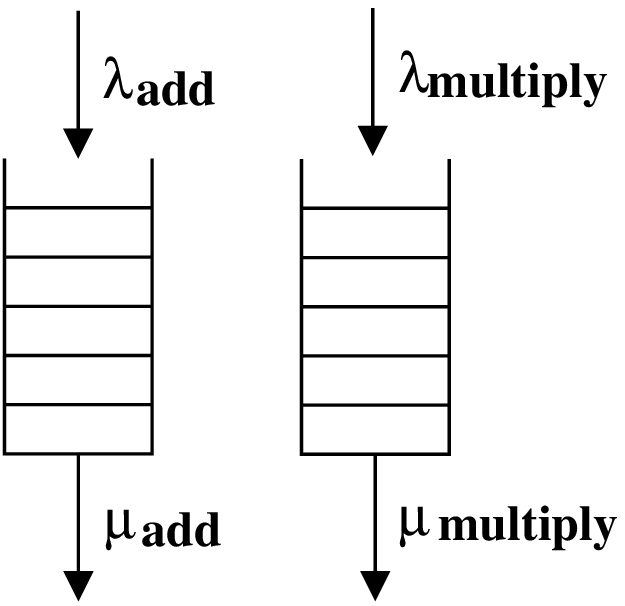}
		\caption{Further abstraction of Fig. \ref{fig:iq_intro:abstract1}}
		\label{fig:iq_intro:abstract1}
	\end{subfigure}
	\qquad \qquad
	\begin{subfigure}{0.45\textwidth}
		\centering
		\vspace{1em}
		\includegraphics[width=0.7\textwidth]{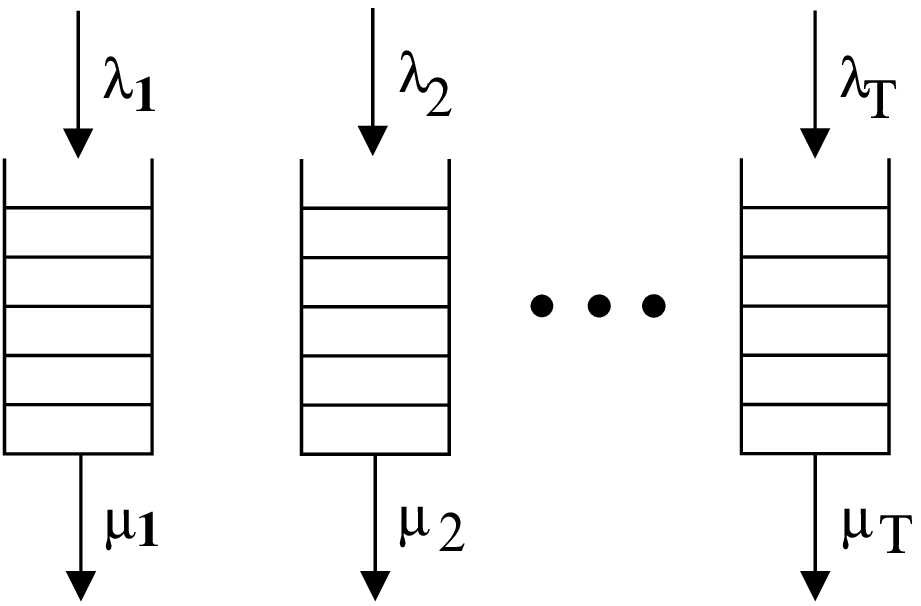}
		\vspace{1em}
		\caption{Generalization of an IQ}
		\label{fig:iq_intro:abstract2}
	\end{subfigure}
	\caption{Fully generalizing an IQ into a framework suitable for a queuing model}
\end{figure}

Abstracting the model even further, we may derive a queuing
network as shown in Figure~\ref{fig:iq_intro:abstract1}.
Here, we have a queuing network with dynamically-sized queues with
the restriction that their total size is less than $N$ (the
physical IQ size) and that the bandwidth does not exceed that of
the physical IQ. In keeping with queuing-theory convention, we
label the distributions of incoming instructions as $\lambda_t$ and FUs
(service units) as $\mu_t$ for each unique instruction type $t$.

Finally, we can generalize the IQ into the queuing network shown
in Figure~\ref{fig:iq_intro:abstract2}.
This generalization depicts a queuing network for an IQ in a
system that supports $T$ unique instruction types, hence we have
$T$ unique queuing lanes. To keep the model consistent, we must
apply to Figure~\ref{fig:iq_intro:abstract2} the constraints that
each logical queue can be up to length $N$, but also that the sum
of all queues cannot exceed $N$.

Given that we are able to model an IQ as a queuing network, we can
model its transitions as a Markov chain. A Markov chain describes
a sequence of events in which the probability of the next state is
determined solely by the current state. In the case of an IQ, we
may consider a Markov chain whose state is the combination of
instructions currently in the IQ. For example, we might say that
the IQ shown in Figures~\ref{fig:iq_intro:basic} and
\ref{fig:iq_intro:equiv} is a Markov chain in state $\langle
4,2\rangle$ since there are four addition and two multiplication
instructions in the IQ.

A compact way to represent a Markov Chain is with a transition matrix $|P|$ with dimension $|S|\times |S|$ where $S$ is the Markov Chain's state space, and each entry $P_{i,j}$ represents the transition probability of the Markov Chain from state $i$ to state $j$ in one time step. E.g., for a Markov chain with $|S|=3$, we may have

\begin{equation} \
	P = \left[ 
		\begin{array}{ccc}
		0.25 & 0.75 & 0.00 \\
		0.30 & 0.60 & 0.10 \\
		0.00 & 0.80 & 0.20
		\end{array} 
	\right] 
	\label{eq:P1}
\end{equation}

In addition to being concise, we can extract from a transition
matrix important properties of the underlying Markov chain. Of
interest to us in this paper is the \emph{steady-state
	distribution}, which, for some transition matrix $P$, is a row vector $\boldsymbol{\pi}$ which exhibits the
property
\begin{equation} 
	\boldsymbol{\pi}P = \boldsymbol{\pi} 
	\label{eq:pi}
\end{equation}
and, more importantly, each element $\pi_i$ holds the steady-state
probability of state $i$, i.e., the percentage of time that the
Markov chain will be in state $i$ as it transitions indefinitely.
From \ref{eq:pi}, it is clear that $\boldsymbol{\pi}$ is obtained by finding the eigenvector of matrix $P$ whose
eigenvalue is equal to 1.0. Finding the steady-state distribution for $P$ from Equation~\ref{eq:P1} leads to
\begin{equation}
\boldsymbol{\pi} = [0.262~~ 0.656~~ 0.082] \label{eq:pi-result}
\end{equation}
According to this distribution, this Markov chain will spend
26.2\% of its time in state 1, 65.6\% of its time in state 2, and
8.2\% of its time in state 3. Intuitively, we may examine $P$ and see that it appears this Markov chain
is usually in state 2, sometimes in state 1, and rarely in state
3, which is in agreement with $\boldsymbol{\pi}$.  If the Markov
chain as in Figure~\ref{eq:P1} is used to
represent the number of instructions waiting in a queue for a
single-instruction-type system, then the steady state result in
Equation~\ref{eq:pi-result} leads to another measurement of
significance, the expected queue length $L$, which is
\[ L = \sum_i i \cdot \pi_i = 0.82 \]

Given that we may derive a transition matrix $P$ for an IQ
modelled as a queuing network and then extract the steady-state
distribution $\boldsymbol{\pi}$, we then know how often the IQ
will be in each state. Recall that a state of the IQ is a
combination of instructions of each type. Thus, extracting
$\boldsymbol{\pi}$ tells us how often there will be, e.g., 4
addition and 2 multiplication instructions in the IQ, as in
Figure~\ref{fig:iq_intro:basic}. Given this distribution, we can
examine how often the IQ will be in an undesirable state, e.g.,
when it is inundated with instructions of a single type and likely
creating a bottleneck in the pipeline. Such a situation may imply
that the FU configuration is unbalanced. Additionally, if the
steady-state distribution reveals that the IQ is in a full state
at a higher than desirable rate, it would be clear that there are
not enough FUs to service the instructions. In this paper we develop the theory to model a CPU's FU-IQ configuration as a novel queuing network and extract the steady state distribution.

This paper is structured as follows. Section~\ref{sec:relatedWork}
discusses some work related to this area, including FU
configuration optimization and queuing networks.
Section~\ref{sec:constructingTheModel} describes the general
algorithm which will be used in this paper to model an IQ as a
Markov chain. Section~\ref{sec:single-analysis} details the
algorithm while considering a processor which supports a single
instruction type, and shows a small example as we proceed through
the section. Section~\ref{sec:generalized-analysis} details the
a processor which supports an arbitrary number of
instruction types, and thus the ultimate goal of this paper.
Section~\ref{sec:example} shows an example of modeling a small IQ
with the queuing network and algorithms described in this paper.
Section~\ref{sec:results} compares the algorithm and model to a
simulation of a benchmark program to validate the model. Section \ref{sec:optimization_problem} uses the presented theory to develop an optimization problem that shows how the proposed model can be applied.

\section{Related Work}
\label{sec:relatedWork} As discussed in Section~\ref{sec:intro},
one could solve the problem of deriving an appropriate FU
configuration by simply having many of each type of FU. However,
the economic drawbacks of such a scheme typically outweigh the
performance gain. One such economic problem is the power
consumption of excess FUs, which must consume power even when not
in use via static power dissipation. This problem was approached
by Rele et al.~\cite{shutdownFUssavepower} who described dynamic
techniques to shut down FUs by up to 90\% of the time with minimal
performance degradation.

Previous research has been done to empirically derive optimal FU
configurations via simulation~\cite{empiricleConfig}. Such
approaches typically only examine the overall throughput of the
processor to determine how well a particular configuration fits a
processor, whereas our goal is to examine the pipeline effects of
each configuration.

Reconfigurable computing~\cite{reconfig1, reconfig2,
	reconfig3_merging_efficiency, reconfig4, reconfig5_datacenters,
	reconfig6_power_efficiency, reconfig7, reconfig8_enhancements,
	reconfig9} aims to do away with pre-configured FU pools. In such a
scheme, an FPGA-like architecture is customized to the application
at hand, and can be altered whenever the instructions to be
executed change. For example, if a reconfigurable computing scheme
was used to run a program consisting of primarily integer
instructions, the majority of the FU-dedicated hardware can be
configured to perform integer operations, while only a small
portion be configured for floating-point operation. However, if
this same architecture were used for a floating-point-intensive
program, it can be configured in the opposite manner.
Reconfigurable computing has been shown to have great increases in
performance and efficiency. However, in such a system, there may
not be \emph{a priori} information ready for reconfiguration in
real time and the process overhead of reconfiguration may easily
outweigh the gain in performance thus acquired.  In addition, the
extra hardware overhead required for the system to be
reconfigurable may easily diminish its intended benefit in cost
efficiency, and the system thus configured may not be at all as
efficient compared to an ordinary system.

None of these approaches mentioned above attempts to derive the
mathematical abstraction of the FU configuration. Having a way to
model the IQ abstractly will allow the derivation of results which
are not affected by the parameters or limitations of a simulator.
Having a way to choose an optimal FU configuration may be able to
optimize cost compared to simpler approaches, and may be able to
optimize performance compared to approaches such as reconfigurable
computing.

Related work in queuing theory describes the solution of many
types of queues~\cite{papoulis2002probability}. Existing queues
and their solutions include many network configurations, including
\begin{enumerate}
	\item Single-queue, single-server
	\item Single-queue, multiple-servers
	\item Multiple-queues, multiple-servers
\end{enumerate}
and also can include various constraints such as finite or
infinite queuing space, or even no waiting area at all (if no
server is ready, an arriving customer leaves immediately).
Impatient customers are also considered, which models customers
that leave a queue after a certain wait time. Networks of queues,
such as a sequence of queues that a customer may traverse, are
also common problems and have many applications in computer
networks including job scheduling and thread
I/O~\cite{mccool2003probability}. Typical parameters to current
queuing networks include the input process (arrival of customers),
the service process (service time distribution), and the number of
servers~\cite{papoulis2002probability}. The typical objectives of
a queuing network problem is to determine the wait times for
customers and the expected queue length.

However, no documented queuing network is applicable to our goal
of modeling a superscalar processor's IQ. Our network must model
a single, finite waiting area. Some unique attributes also apply
to our network:
\begin{enumerate}
	\item Customers in the waiting area wait for a specific type
	of server to be ready
	\item Customers in the waiting area are not necessarily ready
	for service
	\item The arrival and service of customers occur during discrete,
	alternating time periods
\end{enumerate}
The first unique constraint considers that an operation of a
certain type must wait for a FU of that type to be ready. The
second unique constraint models the fact that instructions can be
dispatched to the queue before their operands are ready, rendering
them unready for computation despite occupying a slot in the
finite waiting area. The third considers that a superscalar
processor has discrete pipeline stages: the IQ is partially
emptied as instructions are sent to the FUs, then afterwards the
IQ is replenished with new instructions. This is in contrast to
typical queues, where service events are independent of arrival
events. These properties separate our work from existing queuing
theory work.

\section{Constructing the Model}
\label{sec:constructingTheModel}

In common queuing networks, a closed-form solution can usually be
derived to describe the metrics of the
network~\cite{papoulis2002probability}, including estimated wait
time, estimated queue length, etc. However, a queuing network such
as in Figure~\ref{fig:iq_intro:abstract2} has no known derivation. We aim to develop a model that takes as input two variables:
\begin{enumerate}
	\item The distribution of incoming instruction rates, $\boldsymbol{\lambda}$
	\item The probability of readiness for each instruction types, $\boldsymbol{\rho}$
\end{enumerate}
where elements $\lambda_t \in \boldsymbol{\lambda}$ and $\rho_t \in \boldsymbol{\rho}$ are the incoming distribution and readiness rate for instruction type $t$, respectively. The model is parameterized by the FU-IQ configuration, and produces as output the steady state distribution.
%To model the network without such a closed-form solution, we build
%the model of an IQ as shown in Figure~\ref{fig:per_cc_blackbox}
%which takes the three required information and delivers the
%intended result of IQ usage.
%\begin{figure}[htbp]
%	\centering
%	\includegraphics[width=0.35\textwidth]{per_cc_blackbox.eps}
%	\caption{An Abstraction of the Proposed Queuing Model}
%	\label{fig:per_cc_blackbox}
%\end{figure}
The model is further composed of two sub-models: a consumption model
and an arrival model, with the queuing process modeling each of the
CPU clock cycles as a combination of two discrete steps: instructions in IQ
consumed by the FUs followed by new instructions arriving and
dispatched into the IQ.

Our approach will follow three primary steps and then extract the
target result in a fourth step:
\begin{enumerate}
	\item Construct a transition matrix $C$ to model the
	\emph{consumption} of instructions from the IQ to the FUs during
	the \textbf{issue} stage for each clock cycle
	\item Construct a transition matrix $A$ to model the \emph{arrival}
	of new instructions intro the IQ during the \textbf{dispatch} stage
	for each clock cycle
	\item Show that the transition matrix on a per-clock-cycle
	basis is $P = C\times A$, or the matrix product of the arrival
	and consumption matrices
	\item Extract from $P$ the steady-state distribution
	$\boldsymbol{\pi}$, which tells us the average queue length (IQ occupancy)
\end{enumerate}

Moreover, we are interested in not just the overall IQ occupancy,
but the length of each subqueue (see
Figure~\ref{fig:iq_intro:abstract2}). We set up the model such
that each state contains information about each type of
instruction in the IQ, and therefore reveals the length of each
subqueue for every instruction type.

For the sake of simplicity, we start by considering an IQ in a
system which has just one instruction type and derive the
transition matrix of the IQ. Building on the model of the single
instruction type, we then derive a model for a system with an
arbitrary number of instructions in the same manner. As it turns
out, constructing a model for an arbitrary number of instructions
can be done simply by using the joint probabilities that are derived for
the model with just a single instruction type.

\section{Modelling a Single Instruction Type}
\label{sec:single-analysis}

We begin by considering a system which has only one type of
instruction and, therefore, one type of FU.
%Without loss of
%generality, we may refer to the system's sole instruction type as
%addition.
To model this system as a Markov chain, we first describe that
state space. Since the state space of an IQ Markov chain model is
the set of all possible combinations of IQ occupancy, the state
space for a system with a single instruction type can be described
as
\begin{gather}
\mathcal{S}=\left\{s_i|i\in \mathbb{Z} \cap [0,N]\right\}
\end{gather}
where $N$ is the size of the IQ and $s_i$ represents the state
when the IQ has exactly $i$ instructions currently residing in it.
That is, the state space consists of one state for each possible
number of instructions up to $N$, the size of the IQ. %A graphical representation of the state space of a single instruction type is shown in Figure \ref{fig:example_state_space_1d}. 

To complete the Markov chain model, the state space must be
augmented with transition probabilities between each pair of
states. The transitions we are interested in are the changes in
the number of instructions in the IQ after each clock cycle. In
each clock cycle of the system, two events affect the number of
instructions in the IQ:
\begin{enumerate}
	\item Issuing to the FUs
	\item Dispatching to the IQ
\end{enumerate}

For each of these two events, we derive one transition matrix. We
first derive a transition matrix $C$ which represents the
consumption of instructions out of the IQ and into the FUs in any
clock cycle. We then derive a transition matrix $A$ which
represents the flow of instructions into the IQ in any clock
cycle. We then show that by multiplying these matrices together,
we may derive the complete transition matrix of the IQ; that is,
$P=C\times A$.

\subsection{Single-Type Consumption Model}
\label{sec:consumption_model} In this section, we build the
consumption matrix $C$ to represent the consumption of
instructions from the IQ to the FUs.
During the issue stage of the
pipeline, instructions are sent from the IQ to appropriate FUs.
There are two factors which limit the number of instructions
issued:
\begin{enumerate}
	\item The number of ready instructions in the IQ
	\item The number of available FUs
\end{enumerate}

Instructions may be dispatched into the IQ before their operands
are ready, but instructions may not be issued to FUs until all
operands are ready. During the time an instruction is dispatched
with operands not ready, it cannot be considered ready for issue
but will continue to occupy an IQ entry until it is. Therefore, in
any state the number of instructions which are candidates for
issue is probabilistically distributed based on how many we expect
to have ready operands. %Using these facts, building a model for
%the issue stage is nearly the inverse of building a model for the
%dispatch stage.

Consider transitioning from state $s_i$ to state $s_j$. First, if
$j>i$ we should expect that the transition probability to be zero
during the issue stage of the pipeline since we cannot issue a
negative number of instructions. Therefore, we have
\begin{gather*} p_{s_i,s_j}=0~~~\mbox{if}~~~ i-j<0
\end{gather*}
where $p_{s_i,s_j}$ denotes the corresponding transition
probability.

Second, it is clear that we cannot issue more instructions that we
have FUs available since each FU can process at most one
instruction at a time. Let $\mathcal{F}$ denote the number of FUs
available. Then we have
\begin{gather*}
p_{s_i,s_j}=0 ~~~\mbox{if}~~~ i-j>\mathcal{F}
\end{gather*}

Third, we consider the case that $i-j=\mathcal{F}$, or that every
FU gets issued an instruction. This is simply the case that
\emph{at least} $\mathcal{F}$ instructions in the IQ have their
operands ready. Let $\rho$ represent the probability of an
instruction being ready for issue. Given that there are currently $i$
instructions in the IQ, the transition probability under this
situation is the probability when at least $\mathcal{F}$ are
ready, which can be calculated as
\begin{gather*}
p_{s_i,s_j}=\sum\limits_{k=\mathcal{F}}^{i}\binom{i}{k}\rho^k(1-\rho)^{i-k}
~~~\mbox{if}~~~ i-j=\mathcal{F}
\end{gather*}

Fourth, and finally, we consider the case that $0 \leq
i-j<\mathcal{F}$, or that fewer instructions were issued than
there are FUs. This case reduces to the probability that
\emph{exactly} $i-j$ instructions in the IQ are ready for issue
given that there are $i$ instructions in the IQ. Therefore we have
the corresponding transition probability as
\begin{gather*}
p_{s_i,s_j}=\binom{i}{i-j}\rho^{i-j}(1-\rho)^j ~~~\mbox{if}~~~ 0
\leq i-j<\mathcal{F}
\end{gather*}
Note that for $j=0$, i.e., when every instruction in the IQ was
issued, this equation reduces to
\begin{align*}
p_{s_i,s_j}&=\binom{i}{i-j}\rho^{i-j}(1-\rho)^j \\
&=\binom{i}{i-0}\rho^{i-0}(1-\rho)^0 \\
&=\rho^{i}
\end{align*}
which represents the probability of every instruction in the IQ
being ready for issue.

To summarize the transition probabilities from state $s_i$ to
state $s_j$ during issue, we have
\begin{gather}
\label{eq:single-consumption}
p_{s_i,s_j}=
\begin{cases}
0 & i-j<0 \\
\binom{i}{i-j}\rho^{i-j}(1-\rho)^j & 0 \le i-j<\mathcal{F} \\
\sum\limits_{k=\mathcal{F}}^{i}\binom{i}{k}\rho^k(1-\rho)^{i-k} & i-j=\mathcal{F} \\
0 & i-j>\mathcal{F}
\end{cases}
\end{gather}
We may then use the cases in Equation~\ref{eq:single-consumption}
to populate an issue-stage transition matrix (\emph{consumption}
matrix) for the issue Markov model of an IQ.

We conclude this subsection with a small example. Suppose that we
have a system with an IQ of three entries and two FUs serving
instruction of only one instruction type which has its readiness
probability $\rho$ equal to 0.6. Again with the state space
$\mathcal{S} = \{s_0,s_1,s_2,s_3\}$, the corresponding consumption
matrix of dimension $|\mathcal{S}|\times |\mathcal{S}|$ can be
constructed and populated with
Equation~\ref{eq:single-consumption}:
\begin{equation}
C=
\begin{bmatrix}
1.000 & 0.000 & 0.000 & 0.000 \\
0.600 & 0.400 & 0.000 & 0.000 \\
0.360 & 0.480 & 0.160 & 0.000 \\
0.000 & 0.648 & 0.288 & 0.064
\end{bmatrix}
\end{equation}
in which $C_{i,j}$ denotes the transitioning probability from $i$
instructions to $j$ instructions in IQ. Note that this matrix is
necessarily lower-triangular, since we are describing the
probabilities of instructions being consumed and therefore we must
transition to a state no greater than the current one. Also
interesting to note is that in the lower-left triangle of this
matrix we may also get zero-valued entries. These entries
represent states which cannot be transitioned to due to
insufficient functional units. For example, $C_{3,0}=0$ because
$3-0 > 2$. That is, since we have only 2 FUs, it is impossible to
issue three instructions in one clock cycle (and thus, it is
impossible to transition from state $s_3$ to state $s_0$ in one
issue cycle).

\subsection{Single-Type Arrival Model}
\label{sec:arrival_model} In this section, we derive the
transition matrix $A$ to represent the arrival of new instructions
into the IQ. During the dispatch stage of the pipeline,
instructions are sent from the ROB into the IQ. The number of
instructions which are dispatched depends primarily on two
factors: the number of instructions in the ROB awaiting dispatch
and the number of empty IQ entries. Suppose that the number of
instructions in the ROB in a random clock cycle (i.e., the arrival
rate of instructions) follows some probability mass function (pmf)
$a(x)$. Neglecting the effect of limited bandwidth and given
$a(x)$, we may build an instruction arrival model for any state of
the IQ.

Consider a system in state $s_i$. For any state $s_j$, we shall
build a Markov chain to model the probability of transitioning
from state $s_i$ to state $s_j$ during the dispatch stage of the
pipeline by partitioning the range of values that $i$ and $j$ may
take on. First, we have that the IQ cannot lose instructions
during the dispatch stage (at least 0 must arrive). Therefore we
have
\begin{gather*}
p_{s_i,s_j}=0 ~~~\mbox{if}~~~ j<i
\end{gather*}

Second, consider the case that $i\le j < N$. This transition
implies that $j-i$ instructions have entered the IQ, but that the
IQ is not yet completely full. Therefore, in this case, the
probability of transitioning to state $j$ is the case that
\emph{exactly} $j-i$ instructions are ready for dispatched. Since
the arrival of instructions is modeled by $a(x)$, we have
\begin{gather*} p_{s_i,s_j}=a(j-i) ~~~\mbox{if}~~~ i
\leq j < N
\end{gather*}

Third, and finally, we have the case that $j=N$. In this case,
transitioning from state $s_i$ to state $s_j$ implies that enough
instructions have arrived to fill up the IQ. 
Since this is the last case in a partition of the probability space, the likelihood of this case is the complement of the total likelihood of the previous cases. That is,
\begin{gather}
p_{s_i, s_j} = p_{s_i, s_N} = 1-\sum_{\substack{s_k \in \mathcal{S}\\k\ne N}}a(k-i)
\end{gather}
where $a(k-i)$ is defined by the previous cases.

To summarize the transition probabilities from state $s_i$ to
state $s_j$ during dispatch, we have
\begin{gather}
\label{eq:single-arrival}
p_{s_i,s_j}=
\begin{cases}
0 & j < i \\
a(j-i) & i \le j < N \\
1-\sum\limits_{\substack{s_k \in \mathcal{S}\\k\ne N}}a(k-i) & j=N
\end{cases}
\end{gather}

\begin{figure}
	\begin{subfigure}{0.45\textwidth}
		\centering
		\includegraphics[width=\textwidth]{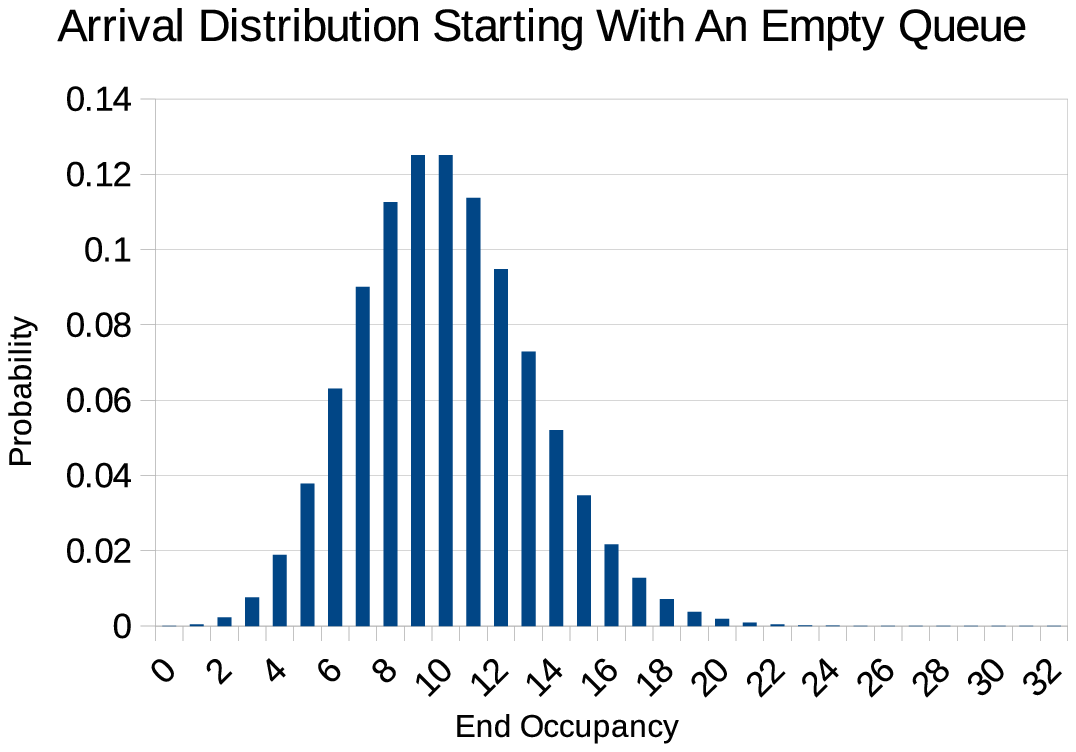}%
		\caption{Empty IQ ($i=0$)}%
		\label{fig:arrival:empty}%
	\end{subfigure}
	\begin{subfigure}{0.45\textwidth}
		\centering
		\includegraphics[width=\textwidth]{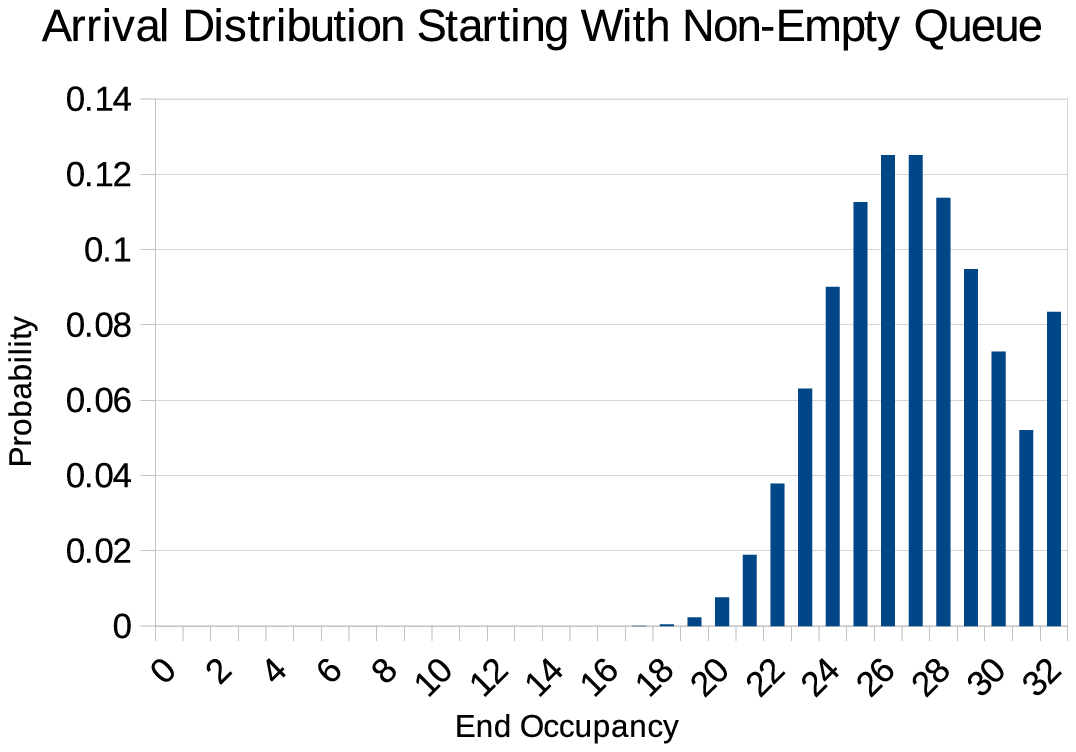}%
		\caption{Occupied IQ ($i=17$)}%
		\label{fig:arrival:non_empty}%
	\end{subfigure}
	\caption{Example arrival distributions, starting from an empty IQ and a partially-occupied IQ, with 32 IQ entries}
\end{figure}

Suppose that the instruction's incoming rate into the IQ follows the
Poisson distribution with mean 1, i.e., 
\begin{gather}
a(k) = \frac{1^ke^{-1}}{k!} = \frac{1}{e k!}
\label{eq:a(x)}
\end{gather}
where $a(k)$ is the probability of $k$ instructions being ready
for dispatch (arrival) to the IQ in any clock cycle. An example of
this equation when applied to a queue of 32 entries is depicted in
Figure~\ref{fig:arrival:empty} for $i=0$ (when the queue is
initially empty).
Another example in Figure~\ref{fig:arrival:non_empty} displays a
case for $i=17$ (when the queue has initially 17 occupied slots).
One can easily see the relationship between the function in
Figure~\ref{fig:arrival:empty} (denoted as $a_0(j)$) and
Figure~\ref{fig:arrival:non_empty} (denoted as $a_{17}(j)$), where
the value of each point in the latter one is exactly shifted 17
positions to the right from the respective point in the former
one; that is,
\[ a_{17}(j) = a_0(j-17) ~~~\mbox{if}~~~ j\geq 17 \]
except for the last point which corresponds to the summation of
all the rest; that is,
\[ a_{17}(32) = \sum_{j=15}^{32} a_0(j) \]
Or, in a generalized form,
\[ a_{i}(N) = \sum_{j=N-i}^{N} a_0(j) \]

We may then use the cases in Equation~\ref{eq:single-arrival} to
populate a dispatch-stage transition matrix (\emph{arrival}
matrix) for the arrival Markov model of an IQ. A small example is
adopted here to illustrate how this matrix is established.  Assume
an IQ with three entries and one instruction type, and the arrival
rate function $a(x)$ follows Equation~\ref{eq:a(x)}. Next, we
consider the state space of this model: $\mathcal{S} = \{\langle 0 \rangle, \langle 1 \rangle, \langle 2 \rangle, \langle 3 \rangle \}$. The arrival matrix of a dimension
$|\mathcal{S}|\times |\mathcal{S}|$ can then be constructed with
one row and column per state. Following
Equation~\ref{eq:single-arrival}, we derive the arrival matrix as
\begin{equation}
A=
\begin{bmatrix}
0.368 & 0.368 & 0.184 & 0.080 \\
0.000 & 0.368 & 0.368 & 0.264 \\
0.000 & 0.000 & 0.368 & 0.632 \\
0.000 & 0.000 & 0.000 & 1.000
\end{bmatrix}
\label{arrival_one:arrival_mat}
\end{equation}
in which $A_{i,j}$ denotes the transition probability from $i$
instructions to $j$ instructions in IQ. 
%Note that this is necessarily an upper-triangular matrix since we have ordered the state indexes such that all preceding indexes represent having fewer instructions in the IQ than the current state. 
%Since we are modeling the arrival, we can only gain instructions from each state. 
From to the ``shift nature'' aforementioned, in the arrival
matrix we see that the arrival distribution in each row is exactly
one entry shifted right from the next row above, except for the
last entry also including the last one from the row above. That is
\begin{eqnarray}
A_{i,j} & = & A_{i-1,j-1}  ~~\forall i,j~~\mbox{s.t.}~~ 1\leq i
\leq j \leq N-1 \\
A_{i,N} & = & A_{i-1,N-1}+A_{i-1,N}
\end{eqnarray}

\subsection{Single-Type Complete model}
\label{sec:complete_model} In Section~\ref{sec:arrival_model} and
\ref{sec:consumption_model}, we partitioned the IQ's behavior
during each clock cycle into two parts: a model $C$ to represent
the consumption of instructions from the IQ into the FUs and a
model $A$ to represent the arrival of new instructions to the IQ.
We now model the change in the IQ between clock cycles by
combining these two models. That is, we use the arrival model and
the consumption model to exhaustively describe the behavior of the
IQ in a single Markov model. Consider an IQ in state $s_i$ at the
beginning of a clock cycle, that is, there are currently $i$
instructions currently in the IQ. During the next clock cycle, the
IQ undergoes two changes: the issue stage and, subsequently, the
dispatch stage. We can therefore denote the change that the IQ
undergoes in one complete clock cycle as
\begin{gather}
\label{eq:complete_stage_transition}
s_i \xrightarrow[]{C} \text{ post-issue stage } \xrightarrow[]{A}
\text{ post-dispatch stage}
\end{gather}
where $\xrightarrow[]{C}$ and $\xrightarrow[]{A}$ represent the
transitions of the issue (consumption of instructions) stage and
dispatch (arrival of instructions) stage, respectively. Now
consider some arbitrary end state $s_j$. To determine the
probability of transitioning from state $s_i$ to state $s_j$
during one clock cycle, we must determine the likelihood that,
after the dispatch stage, the IQ is in state $s_j$. Therefore, we
can rewrite Equation~\ref{eq:complete_stage_transition} as
\begin{gather}
\label{eq:complete_stage_transition_2}
s_i \xrightarrow[]{C} \text{ post-issue stage } \xrightarrow[]{A} s_j
\end{gather}
Furthermore, we note that the post-issue stage is an arbitrary
state and denote it as $s_m$ and write\begin{gather}
\label{eq:complete_stage_transition_3} s_i \xrightarrow[]{C} s_m
\xrightarrow[]{A} s_j
\end{gather}
and we have that the probability of transitioning from state $s_i$
to state $s_j$ in one clock cycle is the probability of
Equation~\ref{eq:complete_stage_transition_3} for all
possibilities of the arbitrary $s_m$. Since each value of $s_m\in
\mathcal{S}$ creates an independent event, we may say that
\begin{align}
p_{s_i,s_j} & = \sum\limits_{s_m \in \mathcal{S}} p\left(s_i \xrightarrow[]{C}
s_m \xrightarrow[]{A} s_j\right) \label{eq:pij_derive:1} \\
& = \sum\limits_{s_m \in \mathcal{S}} p\left(\left(s_i \xrightarrow[]{C} s_m\right)
\wedge \left(s_m \xrightarrow[]{A} s_j\right)\right) \label{eq:pij_derive:2}\\
& = \sum\limits_{s_m \in \mathcal{S}} p\left(s_i \xrightarrow[]{C} s_m\right)
\cdot p \left(s_m \xrightarrow[]{A} s_j\right) \label{eq:pij_derive:3}\\
& = \left(C \times A\right)_{i,j} \label{eq:pij_derive:4}\\
&= P_{i,j} \label{eq:pij_derive:5}
\end{align}
where $C$ is the consumption matrix derived in
Section~\ref{sec:consumption_model} and $A$ is the
arrival matrix derived in
Section~\ref{sec:arrival_model}. Equation~\ref{eq:pij_derive:2} is
derived from Equation~\ref{eq:pij_derive:1} by observing that the
sequence of transitioning can be separated by the discrete nature
of the pipeline. Then we infer Equation~\ref{eq:pij_derive:3} by
noting that the arrival stage is independent of the consumption
stage. Then we observe that by summing over all intermediate
states $s_m$, we are computing the dot product of the $i^{th}$ row
of the consumption matrix and $j^{th}$ column of the arrival
matrix, which corresponds to the value of $P_{i,j}$ where
$P=C\times A$. Therefore, we can build the complete,
per-clock-cycle transition matrix for the IQ by taking the matrix
product of the issue and dispatch matrices.

Continuing with the example shown at the end of
Subsections~\ref{sec:arrival_model} and
\ref{sec:consumption_model} with $\rho=0.6$ and $\lambda = \text{Poi}(1) = \frac{1}{ek!}$, we
conclude this subsection with construction of the per-clock-cycle
transition matrix $P$ from the matrices derived in the preceding
subsections. Thus
\begin{equation}
P=C\times A =
\begin{bmatrix}
0.368 & 0.368 & 0.184 & 0.080 \\
0.221 & 0.368 & 0.258 & 0.154 \\
0.132 & 0.309 & 0.302 & 0.257 \\
0.000 & 0.238 & 0.344 & 0.417
\end{bmatrix}
\end{equation}

Finally, extracting the steady-state distribution  from $|P|$ gives us
\begin{equation}
\boldsymbol{\pi} = [0.171~~ 0.323~~ 0.278~~ 0.231]
\end{equation}
which indicated that this three-entry IQ would be empty during
17.1\% of clock cycles and be full 23.1\% of clock cycles, and the
expected queue length is:
\[ L = \sum_i i \cdot \pi_i = 1.572 \]

\section{Modeling an Arbitrary Number of Instruction Types}
\label{sec:generalized-analysis} In this section, we consider a
system which has an arbitrarily-sized set of unique instruction
types $\{I_1,I_2,...I_T\}$, that is, a system with $T$ different
types of instructions for some $T\in \mathbb{N}$, where each $I_t$
denotes a type of instruction. A system such as this one resembles
a realistic CPU where instructions come in the form in integer
addition, integer multiplication, floating-point addition, etc.,
and each instruction type must be issued to a corresponding type
of FU. We take the simple probability models derived in
Section~\ref{sec:single-analysis} and show that by using joint
probabilities of each instruction type we may develop a model of
FU configuration for an arbitrary system.

\subsection{State Space and State Labeling}
In this subsection we describe the state space and assignment of a
state to an IQ which may contain any number of arbitrary
instruction types. For a system with $T$ instruction types, we can
describe the IQ during an arbitrary clock cycle by the number of
each type of instruction currently inside the IQ. Let $n_t$ denote
the number of instructions of type $I_t$ currently residing in the
queue. We may then view this enumeration as a set of $T$-tuples
$\langle n_1,n_2,...,n_T\rangle $ and use this $T$-tuple as a
state. Collecting all possible $T$-tuples, we have that the state
space of such a system is

\begin{gather*}
\mathcal{S} = \left\{\langle n_1,n_2,...,n_T \rangle  \bigg| n_t
\in \mathbb{N}, \sum_{t}n_t \le N\right\}
\end{gather*}
where $T$ is the number of unique instruction types and $N$ is the
size of the IQ. Lastly, we observe that, with a simple theory of
permutation and combination, the number of states for a system
with $T$ instruction types and an IQ size of $N$ is\begin{gather*}
|\mathcal{S}| = \frac{(T+N)!}{N!T!}
\end{gather*}
which indicates that the state space of this problem is of a size
of exponential of $T$ and $N$, a size simply too large for any
attempt to perform exhaustive simulations to cover them all.

\subsection{Matrix representation}
To create a model for a multi-instruction-type system, we again
will use transition matrices to describe the use of the IQ and its
transitions during both the arrival and consumption stages of the
pipeline. In the previous case of a single instruction type, we
used matrices of size of $|\mathcal{S}|\times |\mathcal{S}|$,
where $\mathcal{S}$ is the state space and $|\mathcal{S}| = N+1$.
Each state $s$ was assigned its own row and column, and the entry
$P_{i,j}$ was the probability of the transition $s_i \rightarrow
s_j$. A similar approach is adopted here for the
multi-instruction-type case, with the only exception being that
states are $N$-tuples rather than integers. That is, for each
model, we will use a transition matrix of size
$|\mathcal{S}|\times |\mathcal{S}|$. Each state $\langle
n_1,n_2,...,n_T\rangle$ will be mapped to an index and assigned
one row and one column in the matrix. Each element of the matrix
$P_{\langle i_1,i_2,...,i_T\rangle,\langle
	j_1,j_2,...,j_T\rangle}$ holds the probability of the transition
$\langle i_1,i_2,...,i_T\rangle \rightarrow \langle
j_1,j_2,...,j_T\rangle$.

\subsection{Multiple-Type Consumption model}
\label{sec:arbitrary_issue_model} In this section, we derive a
model for the issue stage of the pipeline and build a transition
matrix $C$ to describe the probabilities of IQ transition during
the issue stage. In Section~\ref{sec:consumption_model}, we
derived Equation~\ref{eq:single-consumption} to compute the
likelihood of transition of a single type of instruction in the IQ
during the issue stage, i.e., the expected number of instructions
that will be issued in one clock cycle based on the state of the
IQ. We use this equation as a marginal probability and show that
the transition probability in a system with an arbitrary number of
instruction types is the joint probability of all instruction
types as described in Equation~\ref{eq:single-consumption}.

Suppose we have an IQ in some state $s_i = \langle
i_1,i_2,...,i_T\rangle $ and consider some arbitrary state $s_j =
\langle j_1,j_2,...,j_T\rangle$. For the transition $s_i
\xrightarrow[]{C} s_j$ to occur during the issue stage, it must be
the case that, for each instruction type $t$, we issued exactly
$i_t - j_t$ instructions. This case is exactly the joint
probability of the independent events governing the issue of each
instruction type. Therefore, disregarding bandwidth constraints,
we have that
\begin{gather}
\label{eq:arbitrary_consume} p\left( s_i \xrightarrow[]{C} s_j
\right) = \prod_{t=1}^{T} p^{[t]}_{i_t,j_t}
\end{gather}
where $p^{[t]}_{i_t,j_t}$ is defined by
Equation~\ref{eq:single-consumption} for the
single-instruction-type consumption model for instruction type
$I_t$.  Note that each different instruction type may come with a
different readiness parameter $\rho$ for its consumption model.
For example, for a three-instruction-type system with instruction
types from $\{I_1,I_2,I_3\}$, the probability of transitioning
from state $\langle 4,5,3 \rangle$ to state $\langle 3,1,2
\rangle$ is
\[ p(\langle 4,5,3 \rangle \xrightarrow[]{C} \langle 3,1,2
\rangle)
= p^{[1]}_{4,3} \cdot p^{[2]}_{5,1} \cdot p^{[3]}_{3,2}
\]
Represented is matrix form to derive the final consumption matrix
$C$ for this three-instruction-type system, the entry at the row
assigned for state $\langle 4,5,3 \rangle$ and the column assigned
for state $\langle 3,1,2 \rangle$ can then be derived as
\[ C_{\langle 4,5,3 \rangle,\langle 3,1,2 \rangle}
= C^{[1]}_{4,3} \cdot C^{[2]}_{5,1} \cdot C^{[3]}_{3,2}
\]
in which $C^{[t]}_{i_t,j_t}$ represents the respective consumption
matrix for instruction type $I_t$.  Thus
\begin{equation}
C_{s_i,s_j} = \prod_{t=1}^{T} C^{[t]}_{i_t,j_t} \label{eq:multi-C}
\end{equation}

\subsection{Multiple-Type Arrival model}
\label{sec:arbitrary_dispatch_model}

In this section, an arrival model and matrix $A$ for the dispatch
stage of the multi-instruction-type pipeline is to be derived by
using the joint probability of the
Equation~\ref{eq:single-arrival} derived for the
single-instruction-type system.

Similar to state space definition in the consumption model, for
the transition $s_i \xrightarrow[]{A} s_j$ to occur during the
dispatch stage, there are two conditions to consider:
\begin{enumerate}
	\item $\sum\limits_{t} j_t < N$
	\item $\sum\limits_{t} j_t = N$
\end{enumerate}
where $N$ is the size of the IQ. The former case represents when
there is still room in the IQ at the end of dispatch and the model
is quite simple; the latter is a more special case where the IQ
becomes full during dispatch and the model is more complicated, as we
shall see.

\subsubsection{Case 1: $\sum\limits_{t} j_t < N$}

In this case the IQ is not full at the end of the dispatch stage.
This implies all arriving instructions (or all yet-to-dispatch
instructions in the ROB) were able to be dispatched into the IQ, as
evidenced by the leftover space after allocation. Therefore, in
this case, the probability of the transition $s_i
\xrightarrow[]{A} s_j$ is simply the joint probability that, for
each instruction type $t$, exactly $j_t$ instructions were ready
for dispatch. In other words, we have that for $\sum\limits_{t}
j_t < N$
\begin{gather}
\label{eq:arbitrary_arrive_simple} p\left(s_i \xrightarrow[]{A}
s_j \bigg | \sum\limits_t j_t < N \right) = \prod_{t=1}^{T}
p^{[t]}_{i_t,j_t}
\end{gather}
where $ p^{[t]}_{i_t,j_t}$ is defined by
Equation~\ref{eq:single-arrival} for the arrival probability for
the instruction type $I_t$, which is due to that each different
instruction type may come with a different arrival distribution
($\lambda$) for its arrival model.  Similar to the consumption
model derivation process which results in
Equation~\ref{eq:multi-C}, we have for all the entries of this
arrival matrix $A$ that satisfy $\sum\limits_t j_t < N$,
\begin{equation}
A_{s_i,s_j} = \prod_{t=1}^{T} A^{[t]}_{i_t,j_t} \label{eq:multi-A}
\end{equation}

\subsubsection{Case 2: $\sum\limits_{t} j_t = N$}
%/new description
In this second case the IQ becomes full at the end of the dispatch
stage. In the single-type case, this transition is from state $i$
to state $N$ where $N$ is the IQ size. Therefore, this was the
probably that at least $N-i$ instructions were ready to arrive. However, in the
case of multiple instruction types, there are multiple boundary
states. For example, with a 3-entry IQ and 2 instruction types, we
have instead a total of four boundary states $\langle
0,3\rangle,\langle 1,2 \rangle,\langle 2,1 \rangle$ and $\langle
3,0 \rangle$. When the IQ becomes full, we must partition the probability between these boundary states.

Take the example of a transition from state $\langle 0,0 \rangle$
to a boundary state $\langle 1,2 \rangle$. The probability of this
transition are from all events in which at least three
instructions arrive, and that the combination of the first three
instructions dispatched is one of type $I_1$ and two of type
$I_2$.

We can compute the probability of at least three instructions
arriving by subtracting from 1 the probability that fewer than
three instructions arrive using previously-derived equations. To
do so, we can simply iterate over the states whose IQ occupancy is less
than $N$ (which is 3 in this case) and sum all their
probabilities. If we denote the probability of at least three
instructions arriving as $p(3^+)$ and have $\mathcal{S}^*$ denote
the set of all states that are not boundary states, we
have that
\begin{equation}
p(3^+) = 1-\sum_{s_k \in \mathcal{S}^*} p\left(\langle 0,0 \rangle
\xrightarrow[]{A} s_k \right)  \label{eq:p3+}
\end{equation}
Note that each of the term $p\left(\langle 0,0 \rangle
\xrightarrow[]{A} s_k \right)$ in Equation~\ref{eq:p3+} can be
easily obtained by using the formula in
Equation~\ref{eq:arbitrary_arrive_simple} since each state $s_k$
satisfy the condition of case 1.

$p(3^+)$ thus derived includes all the probabilities reaching each
of the four boundary states.  Another probability needs to be
factored in to produce the intended probability of reaching only
state $\langle 1,2 \rangle$.  That is, one of the first three
incoming instructions have to be type $I_1$, and the other two
type $I_2$.  Since order of the three does not matter, the total
number of possible combination is $\frac{(1+2)!}{1!2!}=3$, each
with a probability of
\[ p_{I_1} \cdot p_{I_2}^2 \]
where $p_{I_1}$ and $p_{I_2}$ each denotes the probability that
an arbitrary incoming instruction is type $I_1$ and $I_2$,
respectively.  Note that these values can be easily derived from
the instruction-specific arrival distributions $\lambda_1$ and
$\lambda_2$.  Namely,
\[ p_{I_1} = \frac{\mu_1}{\mu_1+\mu_2},  ~~~~
p_{I_2} = \frac{\mu_2}{\mu_1+\mu_2}
\]
where each $\mu_i$ represents the mean of the incoming distribution $\lambda_i$. This can be generalized for a
$T$-instruction-type system, that is, the probability that the
next incoming instruction being type $I_t$ is
\[ p_{I_t} = \frac{\mu_t}{\sum_{k=1}^{T} \mu_k} \]

For this example, the transitioning probability from state
$\langle 0,0 \rangle$ to a boundary state $\langle 1,2 \rangle$ is
then
\begin{gather}
\left [ 1-\sum_{s_x \in \mathcal{S}^*} p\left(\langle 0,0 \rangle
\xrightarrow[]{A} s_x \right) \right] \frac{(1+2)!}{1!2!}
p_{I_1}^1p_{I_2}^2
\end{gather}
Note that the term $\frac{(1+2)!}{1!2!}$ in this equation is the
total number of permutations, which we also could denote more
formally as the multinomial coefficient of a set of cardinality 3
with element multiplicities of 2 and 1. When extended to a general
case, the total number of permutations becomes $(\sum n_t)!/\prod
(n_t!)$ where $n_t$ represents the number of instructions that
should come in during the current arrival stage.

We may then generalize this equation from some starting state $s_i
= \langle i_1,i_2,...,i_T \rangle$ to some boundary state $s_j =
\langle j_1,j_2,...,j_T\rangle $ as
\begin{align}
&p\left(s_i \xrightarrow[]{A} s_j \bigg | \sum\limits_t j_t = N \right) \nonumber \\
&=  \left [ 1-\sum_{s_x \in \mathcal{S}^*}
p\left(s_i \xrightarrow[]{A} s_x \right)
\right ]\frac{\left(\sum n_t\right)!}{\prod (n_t!)}
\left[\prod_{t=1}^{T} p_{I_t}^{n_t}\right]   \label{eq:edge_trans}
\end{align}
where $n_t=j_t-i_t$, that is, the number of instructions of type
$I_t$ that should be dispatched, and $\mathcal{S}^*$ is the set of
all non-boundary states, i.e.,

\[ \mathcal{S}^* = \mathcal{S} \setminus \left\{\langle n_1,n_2,...,n_T
\rangle  \bigg| \sum_{i}n_i = N\right\}
\]
\subsection{Multiple-Type Complete Model}
\label{sec:complete_model_nd}

Once the consumption model and arrive model are both derived, the
complete model for multi-instruction-type case can be easily
obtained following exactly the same process presented in
Section~\ref{sec:complete_model} for the single-type-instruction
case, namely, by multiplying the consumption and arrival matrices
to produce one transition matrix which completely describes the
IQ's behavior.  That is, $P= C \times A$.

\section{Example}
\label{sec:example}

In this section, we walk through an example implementation of the
proposed algorithm using a simple, small system. Suppose that we
have a system with a three-entry IQ ($N=3$) and two instruction
types $\{I_1,I_2\}$ ($T=2$). Furthermore, we may enumerate the
state space as
\begin{gather*}
\mathcal{S} = \Big\{ \langle 0,0 \rangle, \langle 0,1 \rangle,  \langle 0,2 \rangle, \langle 0,3 \rangle, \langle 1,0 \rangle, \langle 1,1 \rangle, \langle 1,2 \rangle, \langle 2,0 \rangle,  \langle 2,1 \rangle, \langle 3,0 \rangle \Big\}
\end{gather*}
where $\langle n_1,n_2 \rangle$ represents the state during which
the IQ holds $n_1$ and $n_2$ instructions of types $I_1$ and
$I_2$, respectively.

As aforementioned, the proposed algorithm requires three inputs to
produce a model with the following assumptions:
\begin{itemize}
	\item the incoming rates of each instruction type: $\lambda_1 = \text{Poi}(1.5), \lambda_2 = \text{Poi}(1.0)$
	\item the probability of instructions being ready for issue:
	$\rho_1=0.75$ for $I_1$ and $\rho_2=0.8$ for $I_2$;
	\item the FU configuration: two FUs for $I_1$ and one FU for $I_2$.
\end{itemize}

To simplify the state labeling, and enable the mapping of states
to matrix entries, we define a  bijective mapping from the states of the model to a subset of integers as
\[
M = \bigl(\begin{smallmatrix}
\langle 0,0 \rangle & \langle 0,1 \rangle &  \langle 0,2 \rangle & \langle 0,3 \rangle & \langle 1,0 \rangle & \langle 1,1 \rangle & \langle 1,2 \rangle & \langle 2,0 \rangle &  \langle 2,1 \rangle & \langle 3,0 \rangle \\
0 & 1 & 2 & 3 & 4 & 5 & 6 & 7 & 8 & 9
\end{smallmatrix}\bigr)
\]
For each of subsequently-generated consumption, arrival and the
final transition matrices, $C$, $A$, and $P$, it should be
understood that each matrix is of size $10 \times  10$ with their
row and column indices ranging from 0 to 9.  Specifically, the
entry $P_{i,j}$ represents the transition probability from state
$i$ to state $j$ as mapped under this enumeration. For example,
the entry $P_{3,6}$ will hold the transition probability from
state $\langle 0,3 \rangle$ to state $\langle 1,2 \rangle$.

\begin{figure}
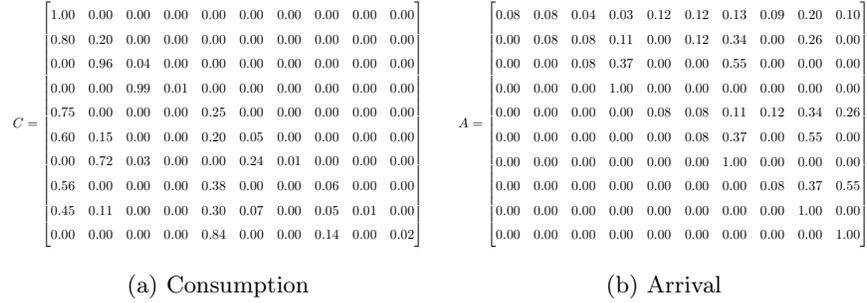

	\centering
	\begin{subfigure}{0.45\textwidth}
		\centering
		\resizebox{\linewidth}{!}{%
			$\displaystyle
			C=
			\begin{bmatrix}
			1.00 & 0.00 & 0.00 & 0.00 & 0.00 & 0.00 & 0.00 & 0.00 & 0.00 & 0.00\\
			0.80 & 0.20 & 0.00 & 0.00 & 0.00 & 0.00 & 0.00 & 0.00 & 0.00 & 0.00\\
			0.00 & 0.96 & 0.04 & 0.00 & 0.00 & 0.00 & 0.00 & 0.00 & 0.00 & 0.00\\
			0.00 & 0.00 & 0.99 & 0.01 & 0.00 & 0.00 & 0.00 & 0.00 & 0.00 & 0.00\\
			0.75 & 0.00 & 0.00 & 0.00 & 0.25 & 0.00 & 0.00 & 0.00 & 0.00 & 0.00\\
			0.60 & 0.15 & 0.00 & 0.00 & 0.20 & 0.05 & 0.00 & 0.00 & 0.00 & 0.00\\
			0.00 & 0.72 & 0.03 & 0.00 & 0.00 & 0.24 & 0.01 & 0.00 & 0.00 & 0.00\\
			0.56 & 0.00 & 0.00 & 0.00 & 0.38 & 0.00 & 0.00 & 0.06 & 0.00 & 0.00\\
			0.45 & 0.11 & 0.00 & 0.00 & 0.30 & 0.07 & 0.00 & 0.05 & 0.01 & 0.00\\
			0.00 & 0.00 & 0.00 & 0.00 & 0.84 & 0.00 & 0.00 & 0.14 & 0.00 & 0.02
			\end{bmatrix}
			$}
		%}
		\caption{Consumption}
		\label{fig:example_joint_consume}
	\end{subfigure}
	\quad 
	\begin{subfigure}{0.45\textwidth}
		\centering
		\resizebox{\linewidth}{!}{%
			$\displaystyle
			A =
			\begin{bmatrix}
			0.08 & 0.08 & 0.04 & 0.03 & 0.12 & 0.12 & 0.13 & 0.09 & 0.20 & 0.10 \\
			0.00 & 0.08 & 0.08 & 0.11 & 0.00 & 0.12 & 0.34 & 0.00 & 0.26 & 0.00 \\
			0.00 & 0.00 & 0.08 & 0.37 & 0.00 & 0.00 & 0.55 & 0.00 & 0.00 & 0.00 \\
			0.00 & 0.00 & 0.00 & 1.00 & 0.00 & 0.00 & 0.00 & 0.00 & 0.00 & 0.00 \\
			0.00 & 0.00 & 0.00 & 0.00 & 0.08 & 0.08 & 0.11 & 0.12 & 0.34 & 0.26 \\
			0.00 & 0.00 & 0.00 & 0.00 & 0.00 & 0.08 & 0.37 & 0.00 & 0.55 & 0.00 \\
			0.00 & 0.00 & 0.00 & 0.00 & 0.00 & 0.00 & 1.00 & 0.00 & 0.00 & 0.00 \\
			0.00 & 0.00 & 0.00 & 0.00 & 0.00 & 0.00 & 0.00 & 0.08 & 0.37 & 0.55 \\
			0.00 & 0.00 & 0.00 & 0.00 & 0.00 & 0.00 & 0.00 & 0.00 & 1.00 & 0.00 \\
			0.00 & 0.00 & 0.00 & 0.00 & 0.00 & 0.00 & 0.00 & 0.00 & 0.00 & 1.00
			\end{bmatrix}
			$}
		%}
		\caption{Arrival}
		\label{fig:example_joint_arrival}
	\end{subfigure}
	\caption{Join consumption and arrival matrices}
\end{figure}

The consumption matrix $C$ is built by first finding, for each
instruction type $I_t$, the respective consumption matrix
$C^{[t]}$, and then following Equation~\ref{eq:arbitrary_consume}
and Equation~\ref{eq:multi-C} with a joint combination process.
The joint consumption transition matrix derived is shown in
Figure~\ref{fig:example_joint_consume}.

As aforementioned, construction of the arrival matrix $A$ is
divided to two steps.  For all the entries that belong to Case 1
in which the end state is not one of boundary states, similar to
the consumption model, it is derived by first finding, for each
instruction type $I_t$, the respective consumption matrix
$A^{[t]}$, and then following
Equation~\ref{eq:arbitrary_arrive_simple} and
Equation~\ref{eq:multi-A} with a joint combination process. For
all the other entries (Case 2) that the end state is one of the
boundary states, Equations~\ref{eq:edge_trans} is used. Also note
that the number of instructions for any instruction type cannot
decrease, that is, the corresponding entry will be 0:
\[\exists k (i_k>j_k) \Rightarrow A_{s_i,s_j} = 0\]
%\[ A_{s_i,s_j} = 0 ~~\mbox{if}~~~ \exists k, ~~\mbox{s.t.}~~~
%	i_k>j_k
%	\]
where $s_i = \langle i_1,i_2,...,i_T\rangle$ and $s_j = \langle
j_1,j_2,...,j_T\rangle$. The final arrival transition matrix $A$
is as shown in Figure~\ref{fig:example_joint_arrival}.

Multiplying the consumption and arrival matrices results in the
per-clock-cycle model $P$ as shown in
Figure~\ref{fig:example_complete_matrix}.
\begin{figure}[htbp]
	\centering
	\resizebox{0.6\linewidth}{!}{%
		$\displaystyle
		P=C\times A = 
		\begin{bmatrix}
		0.08 & 0.08 & 0.04 & 0.03 & 0.12 & 0.12 & 0.13 & 0.09 & 0.20 & 0.10 \\
		0.07 & 0.08 & 0.05 & 0.05 & 0.10 & 0.12 & 0.17 & 0.07 & 0.21 & 0.08 \\
		0.00 & 0.08 & 0.08 & 0.12 & 0.00 & 0.12 & 0.35 & 0.00 & 0.25 & 0.00 \\
		0.00 & 0.00 & 0.08 & 0.37 & 0.00 & 0.00 & 0.55 & 0.00 & 0.00 & 0.00 \\
		0.06 & 0.06 & 0.03 & 0.02 & 0.11 & 0.11 & 0.13 & 0.10 & 0.23 & 0.14 \\
		0.05 & 0.06 & 0.04 & 0.03 & 0.09 & 0.11 & 0.17 & 0.08 & 0.25 & 0.11 \\
		0.00 & 0.06 & 0.06 & 0.09 & 0.00 & 0.11 & 0.36 & 0.00 & 0.32 & 0.00 \\
		0.05 & 0.05 & 0.02 & 0.02 & 0.10 & 0.10 & 0.12 & 0.10 & 0.26 & 0.19 \\
		0.04 & 0.05 & 0.03 & 0.03 & 0.08 & 0.10 & 0.16 & 0.08 & 0.29 & 0.15 \\
		0.00 & 0.00 & 0.00 & 0.00 & 0.07 & 0.07 & 0.10 & 0.12 & 0.34 & 0.31
		\end{bmatrix}
		$}
	\caption{Complete per-clock-cycle transition matrix $P$ for the example.}
	\label{fig:example_complete_matrix}
\end{figure}

Finally, we extract the Left Perron vector $\boldsymbol{\pi}$ from
$P$ to find that the steady-state distribution, expressed as a
percentage of clock cycles, is

\[
	\boldsymbol{\pi} =
	\left[\begin{smallmatrix}
	.026 & .047 & .040 & .066 & .058 & .096 & .228 & .060 & .267 & .110
	\end{smallmatrix}\right]
\]
which in turn represents the probability that one of the ten
possible states occurs. Re-inserting the state labels and displaying $\boldsymbol{\pi}$ as a mapping, we get

\[
\boldsymbol{\pi} = \left(\begin{smallmatrix}
\langle 0,0 \rangle & \langle 0,1 \rangle &  \langle 0,2 \rangle & \langle 0,3 \rangle & \langle 1,0 \rangle & \langle 1,1 \rangle & \langle 1,2 \rangle & \langle 2,0 \rangle &  \langle 2,1 \rangle & \langle 3,0 \rangle \\
.026 & .047 & .040 & .066 & .058 & .096 & .228 & .060 & .267 & .110 
\end{smallmatrix}\right)
\]

\begin{figure}[htbp]
	\centering
	\includegraphics[width=3.3in]{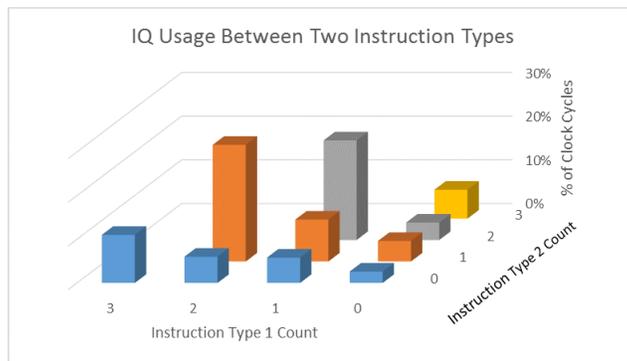}
	\caption{Graphical Representation of The Results}
	\label{fig:2dexample_state_space}
\end{figure}

If we neatly rearrange these probabilities into an organized state
space we obtain the result as shown in
Figure~\ref{fig:2dexample_state_space}. On the horizontal axis is
the number of instructions of type $I_1$, and on the depth axis is
the number of instructions of type $I_2$. Vertically, we show the
estimated percentage of time the IQ will spend in each state.

From the results of the Left Perron vector, we see that the IQ is
full during approximately 67\% of clock cycles, when $n_1+n_2=3$ for
all states $\langle n_1,n_2 \rangle$. One conclusion we can make
is that adding more FUs to the system would alleviate the
bottleneck in the IQ since it is full relatively often. In
addition, the IQ usage is slightly dominated by instructions of
type $I_1$, which is because the arrival rate of
type $I_1$ ($\mu_1=1.5$) is higher than one of type $I_2$
($\mu_2=1$), coupled with the fact that the readiness
parameter for type $I_1$ ($\rho=0.75$) is lower than type $I_2$ ($\rho=0.8$). Furthermore, we may derive the average IQ
usage of each instruction type by summing over the steady state.
For some instruction type $I_t$, we have that $L_t$, the average
number of instructions of type $t$ in the IQ, can be derived as
\begin{gather}
L_t = \sum_{s\in \mathcal{S}}n_t\boldsymbol{\pi}[s]
\label{eq:queue_length_individual}
\end{gather}
where $n_t$ is the number of instruction of type $I_t$ in state
$s$. That is, we sum the $t^{th}$ index of each state $s$
multiplied by the probability of being in state $s$. We have then
\[ L_1=1.366, L_2=1.144 \]
The overall queue length can be derived using
\begin{gather}
L = \sum_{s\in \mathcal{S}} \left(\sum_{t=1}^{T} n_t\right)
\boldsymbol{\pi}[s]
\end{gather}
or simply
\[ L= \sum_{t=1}^{T} L_t \]
which leads to $L=2.51$, a relatively tight IQ utilization.

The most important analysis is to determine if the employment of
various functional units leads to the best throughput. Adding more
FUs for sure will increase throughput, but it remains to be
determined that which type of FU should be invested in order to
obtain the most increase in throughput. Note that for all
instructions of various types to flow through IQ in a
``congruent'' manner, that is, no instructions of any type is more
``stagnant'' than the others, which would subsequently cause
instructions of other types to slow down as well due to data
dependency among instructions across types. Such a congruency can
be measured by how closely matched the ratios of arrival rate and
expected queue length among all instruction types are.  Let this
``flow ratio'' be denoted as $R_t$ for instruction type $I_t$, and
\[  R_t = \frac{\mu_t}{L_t} \]
where $\mu_t$ is the mean of the incoming distribution for type $t$.
The higher $R_t$ is, the better the instructions of type $I_t$ are
flowing through the IQ. Thus we have
\[ R_1=1.098, R_2=0.874 \]
which shows that the FUs for type $I_2$ are more utilized than
those for type $I_1$.  This may call for an increase of the FU
number for $I_2$ to see if the balance between $R_1$ and $R_2$ can
be further improved. Or strictly theoretically speaking, the
number of functional units for type $I_2$ should be increased by a
factor of 1.256 (i.e., $R_1/R_2$) in order to match the flow ratio
between the two types of instructions.

\section{Model Validation}
\label{sec:results}
In this section we apply the proposed technique and validate its accuracy. That is, we intend to experiment with how well the proposed model predicts the IQ usage under a given configuration. We will experiment using the SimpleScalar~\cite{simplescalar} simulator with the SPEC CPU 2006 benchmark suite~\cite{spec2006}. The benchmarks are compiled with the Alpha instruction set architecture. The simulator configuration used in shown in Table \ref{table:sim_config}.

\begin{table}
	\caption{Parameters for the Simulation Environment Used in the Experiments}
	\label{table:sim_config}
	\centering
	\begin{tabular}{ |c|c| }
		\hline
		Parameter & Value \\
		\hline\hline
		Decode/Issue/Commit Width & 8/8/8 \\
		\hline
		LSQ/ROB/IQ/RF/WB Size & 48/128/16/256/16  \\
		\hline
		Mem. Ports/IALU/FP Mult./FP ALU & 2/4/1/1\\
		\hline
	\end{tabular}
\end{table}

\begin{figure}[t]
	\centering
	\includegraphics[width=3.3in]{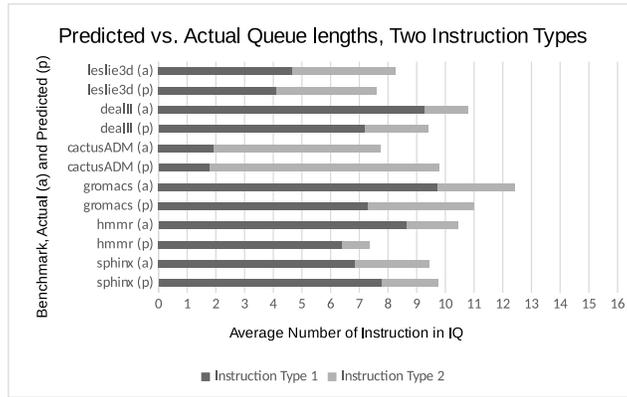}
	\caption{Modeled Predicted Queue Lengths vs. Actual Average
		Queue Lengths for Various SPEC CPU 2006 Floating-point Benchmarks}
	\label{fig:results:all}
\end{figure}

Generally, queuing theory solutions are parameterized and leave it to the application to infer the parameters to use the model. Since the proposed model is a queuing theory model and therefore requires parameters \textit{a priori} (the values for $\lambda$ and $\rho$ ), we validate the model by first running simulations and gathering statistics. We then use the statistics to infer the parameters to the model. We then query the model with the parameters to predict the queue lengths. We then compare the models prediction vs. the simulation results to validate how closely the model matches the IQ-FU instruction flow. We focus on the two most utilized instruction types from each benchmark. We find that in practice in this environment, an individual benchmark uses no more than two instruction types in meaningful quantities. Figure~\ref{fig:results:all} shows the results from the simulations comparing the predicted queue lengths against the actual values. 
Overall, a very respectable average prediction error rate of
20.1\% is observed. The discrepancies most likely are due to the
parameters empirically acquired, namely the $\lambda$'s and
$\rho$'s which may vary from benchmark to benchmark by a somewhat
significant margin.

\section{As An Optimization Problem}
\label{sec:optimization_problem}
In this section we use the proposed model to create an optimization problem that can be applied in practice. Suppose we have some FU configuration $\mathcal{F}$, where $|\mathcal{F}_t|$ represents the number of FUs for instruction type $t$. Using the proposed model, we can determine the mean length of the queue for each instruction type $t$ under the configuration $\mathcal{F}$ as $\ell(t, \mathcal{F})$ via Equation \ref{eq:queue_length_individual}. Since a lower mean queue length indicates faster throughput, $\ell(t, \mathcal{F})$ becomes more costly as it rises. 

Similarly, each FU of type $t$ costs $c(\mathcal{F}_{t})$, and therefore having $|\mathcal{F}_{t}|$ FUs costs $|\mathcal{F}_t|\cdot c(\mathcal{F}_{t})$. Therefore we can say that the total cost of configuration $\mathcal{F}$ is 
\begin{equation}
	C(\mathcal{F}) = \sum\limits_{t} \left( 
	\ell(t, \mathcal{F}) + c(\mathcal{F}_{t})\cdot|\mathcal{F}_{t}| \right)
	\label{eq:cost}
\end{equation}
For $T$ instruction types, equation \ref{eq:cost} results in a convex, $T$-dimensional curve embedded in a $(T+1)$-dimensional space, and minimizing for some given cost parameters would result in an optimal $\mathcal{F}$. 

\begin{figure}[t]
	\begin{subfigure}{0.48\textwidth}
		\centering
		\includegraphics[width=\textwidth]{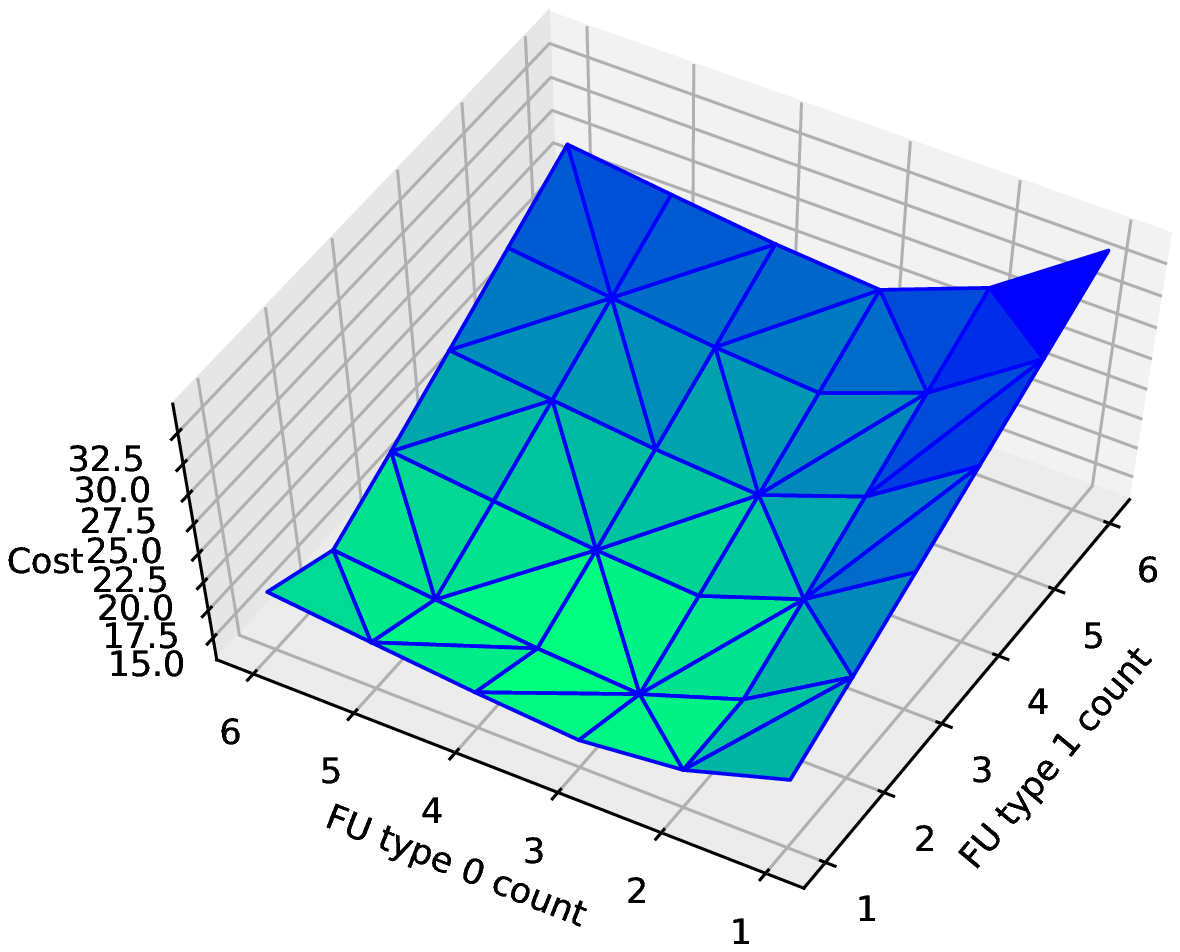}%
		\caption{With FU costs $[1,3]$}%
		\label{fig:costs:1_3}%
	\end{subfigure}
	\begin{subfigure}{0.48\textwidth}
		\centering
		\includegraphics[width=\textwidth]{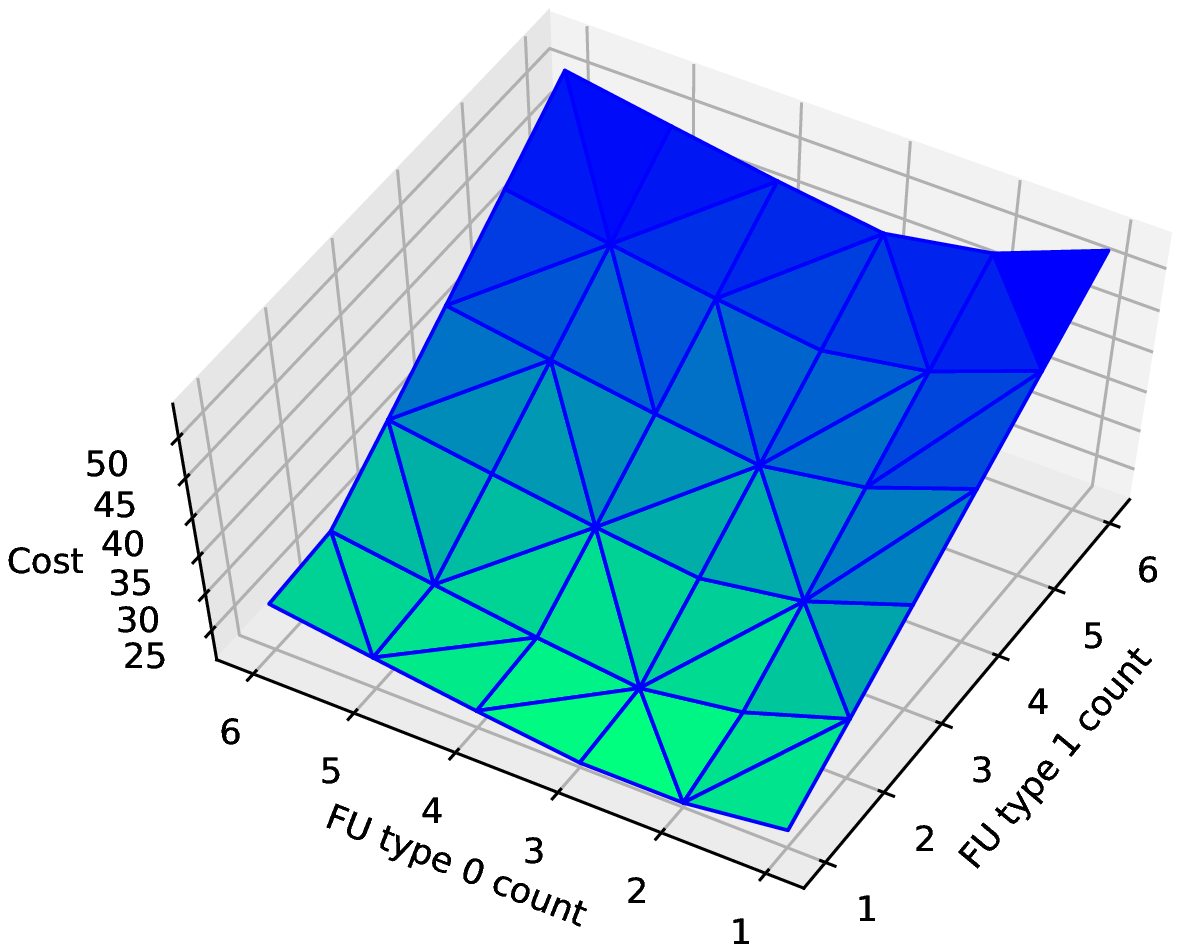}%
		\caption{With FU costs $[2,6]$}%
		\label{fig:costs:2_6}%
	\end{subfigure}
	\caption{Results of creating an optimization problem, where the cost of some FU configuration is the sum of the queue lengths and the cost of the FU configuration}
	\label{eq:costs:main}
\end{figure}

For example, say we have $t_0$ as multiplication and $t_1$ as division. Since dividers require a more complicated circuit and more power than multiplication, we can assign costs $c(\mathcal{F}_0)=1$ and $c(\mathcal{F}_1) = 3$. If we have an IQ of size 16,  incoming distributions $\lambda_0 = \text{Poi}(2), \lambda_1 = \text{Poi}(1)$, and readiness rates $\rho_0 = \rho_1 = 0.8$, we can derive the cost curve across the state space of $\mathcal{F}$. Examining Figure \ref{fig:costs:1_3}, we see that under these parameters the optimal FU configuration is $|\mathcal{F}_0| = 3, |\mathcal{F}_1| = 2$ (i.e., 3 multipliers and 2 dividers) with a cost of $13.4$. If we double the costs of the FUs and fix the remaining parameters, we see in Figure \ref{fig:costs:2_6} that the optimal FU configuration drops to $|\mathcal{F}_0| = 2, |\mathcal{F}_1| = 1$ for a cost of $21.5$, as we should expect since the only parameters to have changed are the costs of the FUs which are now more expensive relative to queue length.

Lastly, note that since \ref{eq:cost} is convex, in practice we can use hill climbing to quickly find an optimal configuration starting from a configuration of 1 FU of each type and ascending through the state space.

\section{Conclusion}
\label{sec:conclusion} In this paper we described a novel queuing network comprised of a shared waiting area with variably-sized pools of unique server types. Customers in the waiting area wait for availability of a specific type of server and cannot be served by any other type. Furthermore, a customer in the waiting area is not necessarily ready for service. We showed how to fit a superscalar processor's IQ-FU configuration to such a model with instructions as customers and functional units as servers. We showed how to solve for the expected queue length of each instruction type, revealing the efficiency of an arbitrary FU configuration and the IQ usage of every instruction type. We also showed how the proposed theory can be applied in practice as an optimization problem.

%\bibliographystyle{unsrt}
%\bibliography{cites}
%\input{elsarticle-template.bbl}

\end{document}